\documentclass[12pt, onecolumn]{IEEEtran}
\usepackage{amsmath,amssymb,euscript,yfonts,psfrag,latexsym,dsfont,graphicx}
\usepackage{bbm,color,amstext,wasysym,subfig,parskip,balance,cite}
\graphicspath{{./},{./figures/}}

\newtheorem{thm}{Theorem}

\newtheorem{problem}{Problem}

\newcommand{\mE}{{\mathbb E}}

\newcommand{\mR}{{\mathbb R}}

\newcommand{\cN}{{\mathcal N}}

\newcommand{\cR}{{\mathcal R}}

\newcommand{\tr}{\operatorname{tr}}

\newcommand{\range}{\cR}

\definecolor{grey}{rgb}{0.6,0.6,0.6}
\definecolor{lightgray}{rgb}{0.97,.99,0.99}

\begin{document}
\title{Covariance steering in zero-sum\\ linear-quadratic two-player differential games\thanks{Partial support was provided by NSF under grants 1665031, 1807664, 1839441 and 1901599, and by AFOSR under grant FA9550-17-1-0435.}}

\author{Yongxin Chen, Tryphon T. Georgiou and Michele Pavon
\thanks{Y.\ Chen is with the School of Aerospace Engineering, Georgia Tech, Atlanta, GA; {email: yongchen@gatech.edu}}
\thanks{T. T. Georgiou is with the Department of Mechanical Engineering, UCI, Irvine, CA; {email: tryphon@uci.edu}}
\thanks{M.\ Pavon is with the Dipartimento di Matematica,
Universit\`a di Padova, via Trieste 63, 35121 Padova, Italy; {email: pavon@math.unipd.it}}}

\maketitle

\begin{abstract}
We formulate a new class of two-person zero-sum differential games, in a stochastic setting, where a specification on a target terminal state distribution is imposed on the players. We address such added specification by introducing incentives to the game that guides the players to steer the join distribution accordingly. In the present paper, we only address linear quadratic games with Gaussian target distribution. The solution is characterized by a coupled Riccati equations system, resembling that in the standard linear quadratic differential games. Indeed, once the incentive function is calculated, our problem reduces to a standard one. Tthe framework developed in this paper extends previous results in covariance control, a fast growing research area. On the numerical side,  problems herein are reformulated as convex-concave minimax problems for which efficient and reliable algorithms are available.
\end{abstract}

\section{Introduction}
Differential games \cite{Isa99,DocJorVanSor00} represent a class of games where players are constrained by continuous-time dynamics. They represent a natural marriage of game theory and optimal control, and have had a huge impact in economics, management sciences, operations research as well as more traditional control involving several interacting agents \cite{Isa99,DocJorVanSor00}. Standard (non-cooperative) differential games stipulate that each player seeks a strategy that maximizes her/his payoff. Thus, the counterpart of optimal control policy in this setting is a Nash equilibrium \cite{Osb04} --the optimal strategy for each requires/assumes that the other agents choose an equilibrium-policy where none gains by moving away from. In general, differential games are challenging both in theory as well as in implementation and numerics.

In this work, we consider two-player zero-sum differential games, a special type of differential games involving two agents competing with each other where one's gain is precisely the other's loss. Such models are often encountered in economics \cite{DocJorVanSor00}, and turned out to play an important role in robust control theory \cite{BasOls99}. From an optimization point of view, deriving a Nash equilibrium strategy is equivalent to finding a saddle point of the utility function. For general minimax optimization \cite{Ber97}, a global, sometimes even local, saddle point may not exist. One condition that guarantees existence of global saddle point is that the cost is convex over the minimization variable and concave over the maximization variable \cite{NedOzd09,CheLanOuy14}. In this paper, we focus on linear-quadratic games \cite{LukRus71,Bar76,Ber79,Zha05,Del07} where the convex-concave assumption is indeed valid. 

In most of the literature on differential games, the focus is on existence and properties of solutions to specific problems. Herein, we advance a somewhat different perspective, and seek a systematic approach of introducing incentives to the game, so as to steer the response of the players towards some desirable pattern. In general terms, our rationale is similar to
that of ``mechanism design,'' in economic theories, to regulate via incentives.  More precisely,
it is akin to a cost selection in optimal control problems, so as to induce an optimal policy with desired properties. However, one major difference is that in control problems, there is a direct correspondence between cost and optimal policy, whereas in differential games, in principle, the agents may respond in a variety of ways to the same incentive. The situation is
remedied and the correspondence is restored when we hypothesize that the players are rational and follow Nash equilibrium policies. 

In the present work, specifically, we consider stochastic games and our goal is to design a proper incentive so that the system state of the players reaches a target probability distribution, always under the tacit assumption that they follow a Nash policy. In other words, although the game is noncooperative in that each player seeks maximal advantage, an implicit level of cooperation is imposed by the end-point requirement of specifying a target state distribution. 
Thus, while in economics and social sciences, game theory provides a model and an explanatory frame \cite{friedman1990game}, our setting is motivated by design where we seek to modify the players behavior in a manner that forces the said terminal target distribution.
The linear-quadratic setting is amenable to develop the basic paradigm of how to steer antagonistic players in a zero-sum game towards a desirable stochastic state distribution via a suitable choice of a quadratic teminal cost to serve as an incentive. We note that a similar viewpoint has been explored in a mean-field-game setting,  representing competition among a huge group of identical agents, with the goal to steer the distribution of the cohort via incentives \cite{CheGeoPav18}.

The present builds on recently developed theory of optimal control with stochastic state-constraints that is known (in the linear-quadratic Gaussian setting) as covariance control \cite{HotSke87,skelton1989covariance,yasuda1993covariance,skelton1997unified,CheGeoPav14a,CheGeoPav14b,CheGeoPav14c,CheGeoPav17a,Bak16,RidTsi18a,OkaTsi19,Bak18}. Covariance control is motivated by the desire to enforce precise probabilistic specifications so as to reduce conservativeness and improve effectiveness of control strategies. 
Surprisingly, covariance control ties stochastic control with two other seemingly disconnected research topics with a long history, {\em optimal mass transport} (OMT) and {\em Schr\"odinger bridges} (SB), see \cite{CheGeoPav16} and the references therein. Despite clear differences in the rationale for covariance control with that of the present paper, there are deeper points of contact and, in fact, the latter can be viewed as an extension of the former. 

The paper is structured as follows. Section \ref{sec:pre} contains background material as well as a formulation of our basic problem. Its solution is given in Section \ref{sec:main} whereas results for the case of infinite horizon are presented in Section \ref{sec:inf}. We conclude with a few short remarks in Section \ref{sec:con}.

\section{Preliminaries and problem formulation}\label{sec:pre}
We briefly introduce the covariance control and linear quadratic differential game theories. Both are needed to solve the covariance steering problems in differential games, whose problem formulation is also presented in this section.

\subsection{Covariance control}\label{sec:cov}
We begin by considering a standard linear stochastic system
	\begin{equation}
		dx(t) = Ax(t) dt + B u(t) dt + Cdw,~x(0)\sim \cN(0,\Sigma_0),
	\end{equation}
where $x \in \mR^n$ denotes the state, $u\in \mR^m$ is the control input, and $w$ is a standard Wiener process (Brownian motion). The pair $(A,\, B)$ is assumed to be controllable.
The goal in covariance control over, e.g., a finite time interval \cite{CheGeoPav14a,CheGeoPav14b}, is to determine an optimal feedback control law that drives the state from an initial Gaussian distribution $\cN(0, \Sigma_0)$ to a target terminal Gaussian distribution $\cN(0, \Sigma_T)$\footnote{For the simplicity of exposition, we focus only on zero-mean distributions. The cases with nonzero-mean distributions can be analyzed similarly.} at time $t=T$ while minimizing the quadratic cost functional
	\begin{equation}
		J (u) = \mE\{\int_0^T[ x(t)'Qx(t) + \|u(t)\|^2 ] dt\}.
	\end{equation}
	As usual, $Q$ is a suitable matrix that defines cost in the state variables.
It is a nonstandard stochastic control problem due to the constraint $x(T)\sim \cN(0,\Sigma_T)$. Nevertheless, the optimal policy is similar to that in standard linear-quadratic regulator theory and given in state feedback form \cite{CheGeoPav14a,CheGeoPav14b},
	\[
		u(t,x) = -B'\Pi(t) x,
	\]
where $\Pi$ is a time-varying matrix which, together with a second time-varying matrix $H$, satisfies the following coupled system of differential equations
	\begin{subequations}\label{eq:sb}
	\begin{eqnarray}
	\dot \Pi &=& -A'\Pi- \Pi A +\Pi BB'\Pi -Q
	\\
	\dot H &=& -A'H-HA-HBB'H+Q
	\\&&+(\Pi+H)(BB'-CC')(\Pi+H) \nonumber
	\\ \Sigma_0^{-1} &=& \Pi(0)+H(0),
	\\ \Sigma_T^{-1} &=& \Pi(T)+H(T).
	\end{eqnarray}
	\end{subequations}

Equations \eqref{eq:sb} relate to a system of nonlinear equations in classical 1932-work by Erwin Schr\"odinger on the so-called Schr\"odinger bridge problem \cite{wakolbinger1992bridges}, and can also be seen to constitute a coupled pair of a Fokker-Planck equation and a Hamilton-Jacobi equation, see \cite{CheGeoPav14b,CheGeoPav16}. When $B=C$, $\Pi$ and $H$ are only coupled through the boundary conditions. In this special case, the equations have a closed-form solution \cite{CheGeoPav14a,CheGeoPav14c}. In general, not only closed-form solutions may not exist but even the existence remains an open research topic  \cite{CheGeoPav14b}. However, a numerical scheme allows constructing solutions that are approximately optimal to an arbitrary precision \cite{CheGeoPav14b} and so, from a practical point of view, the covariance control problem described in this section can be considered completely solved. 
	
\subsection{Linear quadratic differential games}\label{sec:LQDG}
Consider now the stochastic dynamical system 
	\begin{subequations}\label{eq:lqgame}
	\begin{equation}
		dx(t) = Ax(t) dt + B_1 u(t) dt + B_2 v(t) dt + Cdw,~x(0)\sim \rho_0
	\end{equation}
where $x \in \mR^n$ is the (combined) state, $u\in \mR^m, v\in \mR^p$ are the control inputs of the two players, player $1$ and $2$, respectively, and $w$ is standard Wiener process as before. The pairs $(A,\, B_1)$ and $(A,\, B_2)$ are assumed to be controllable. The initial state $x(0)$ is taken to be a random vector with probability distribution $\rho_0$. The two players compete with each other aiming at minimizing their own cost through proper feedback control policies \cite{Isa99}. 

The cost function for agent $1$ is 
	\begin{eqnarray}
		J_1 (u,v) &= & \mE\{\int_0^T [x(t)'Qx(t) + \|u(t)\|^2 - \|v(t)\|^2] dt \nonumber
		\\&&+ x(T)'Fx(T)\}
	\end{eqnarray}
and the cost for agent $2$ is 
	\begin{equation}
		J_2 (u, v) = - J_1(u,v),
	\end{equation}
For suitable quadratic forms specified by matrices $Q$ and $F$ that dictate running and terminal state cost, respectively.
Overall, in this zero-sum-game setting, agents/players seek the solution of the minimax problem
	\begin{equation}
		\min_{u} \max_{v}\, J_1(u, v)
	\end{equation}
	\end{subequations}
where the optimization is taken over all the feasible feedback control policies. For  notational simplicity we take $\|u(t)\|^2$, $\|v(t)\|^2$ as the running cost for the respective control variables, that is, with respective ``weights'' the identity matrices. More generally, when using suitable weighted quadratic norms, one may proceed in a similar manner. Either way, the solution is similar to that in linear quadratic control problems and is provided in the following theorem \cite{LukRus71,Bar76,Ber79}.

\begin{thm}\label{thm:lqgame}
If the solution to the Riccati equation
	\begin{equation}\label{eq:riccati}
		\dot \Pi +A'\Pi+ \Pi A -\Pi (B_1B_1'-B_2B_2')\Pi +Q = 0,\,\Pi (T) = F
	\end{equation}
exists over the time interval $[0,\,T]$, then the differential game associated with \eqref{eq:lqgame} has a unique solution, given by the linear state feedback control 
	\begin{subequations}\label{eq:nash}
	\begin{eqnarray}
	u^*(t,x) &=& -B_1'\Pi (t)x\\
	v^*(t,x) &=& B_2'\Pi (t)x.
	\end{eqnarray}
	\end{subequations}
\end{thm}

We provide an elementary derivation below that relies on ``completion of squares.'' Complete arguments to further argue uniqueness follow along similar lines in \cite{Bar76,Zha05,Del07}, to which we refer for specifics.
Let $\Pi$ be the solution to the Riccati equation \eqref{eq:riccati}, then $\mE\{x(0)'\Pi (0)x(0)\}$ is clearly a constant depending only on the initial distribution $\rho_0$.  
Thus, the cost $J_1(u,v)$ is equivalent to 
	\begin{eqnarray*}
	&&
	\mE\{\int_0^T[ x(t)'Qx(t) + \|u(t)\|^2 - \|v(t)\|^2] dt 
	\\&&+ d[x(t)'\Pi (t)x(t)]\}
	\\&=&\mE\{\int_0^T[ x(t)'Qx(t) + \|u(t)\|^2 - \|v(t)\|^2] dt 
	\\&&+ (Ax(t) dt + B_1 u(t) dt + B_2 v(t) dt + Cdw)'\Pi (t)x(t)
	\\&&+x(t)'\Pi (t)(Ax(t) dt + B_1 u(t) dt + B_2 v(t) dt + Cdw)
	\\&&+\tr(\Pi CC' )dt+x(t)'\dot \Pi (t)x(t)dt\}
	\\&=& \mE\{\int_0^T \|u(t)+B_1'\Pi (t)x(t)\|^2-\|v(t)-B_2'\Pi (t)x(t)\|^2
	\\&& + \tr(\Pi CC')dt\},
	\end{eqnarray*}
where we have used the Riccati equation \eqref{eq:riccati} and the fact that $\mE \{dw\}$ is zero in the above. Clearly, the policy in \eqref{eq:nash} is a stationary point of $J_1(u,v)$. Indeed, for any feedback control policy $u, v$, we have that
\[
	J_1(u^*\!,v^*)\!-\! J_1(u^*\!,v)= \mE\{\int_0^T\!\!\! \|v(t)-B_2'\Pi (t)x(t)\|^2 dt\} \!\ge\! 0,
\]
and
\[	J_1(u,v^*)\!-\! J_1(u^*\!,v^*) = \mE\{\int_0^T\!\!\! \|u(t)-B_2'\Pi (t)x(t)\|^2 dt\}\! \ge\! 0.
\]

\subsection{Problem formulation}
We consider the covariance control problem over a differential game system. Departing from the problem in \eqref{eq:lqgame}, in our new problem the terminal cost $x(T)'Fx(T)$ is a design parameter and the goal is to drive the state to a target terminal distribution. This is formally stated as follows.
\begin{problem}\label{pro:main}
Consider the differential game with stochastic dynamics 
	\begin{eqnarray*}
		dx(t) &=& Ax(t) dt + B_1 u(t) dt + B_2 v(t) dt + Cdw
		\\&&x(0)\sim \cN(0,\Sigma_0)
	\end{eqnarray*}
and cost $J_2 = -J_1$, 
	\begin{eqnarray*}
		J_1 (u,v) &=& \mE\{\int_0^T [x(t)'Qx(t) + \|u(t)\|^2 - \|v(t)\|^2] dt
		\\&& + x(T)'Fx(T)\}.
	\end{eqnarray*}
Here $F$ is a design parameter. Determine a value for $F$ so that the state reaches the specified target distribution $\cN(0, \Sigma_T)$ at terminal time $t=T$, assuming that the two players are rational. 
\end{problem}

We remark that the Riccati equation \eqref{eq:riccati} may fail to have a solution \cite{LukRus71,Del07}. The choice of $F$ in that case requires great care.

\section{Main results}\label{sec:main}
Starting from a proper $F$, the Nash equilibrium is characterized by Theorem \ref{thm:lqgame}, yielding the closed loop system
	\[
		dx = A x -B_1B_1'\Pi x + B_2 B_2'\Pi x + Cdw.
	\]
Its state covariance $\Sigma(t)=\mE\{x(t)x(t)'\}$ satisfies the Lyapunov equation
	\begin{eqnarray}
	\dot \Sigma &=& (A-B_1B_1'\Pi+B_2B_2'\Pi)\Sigma\nonumber
	\\&&+\Sigma(A-B_1B_1'\Pi+B_2B_2'\Pi)'+CC'\label{eq:lyap}
	\end{eqnarray}
with initial condition $\Sigma(0) = \Sigma_0$. 
Let 
	\[
		H = \Sigma^{-1} - \Pi.
	\]
Then, a straightforward calculation shows that $H$ satisfies the differential equation
	\begin{eqnarray*}
		\dot H &=& -A'H-HA-H(B_1B_1'-B_2B_2')H+Q
		\\&&+(\Pi+H)(B_1B_1'-B_2B_2'-CC')(\Pi+H)
	\end{eqnarray*}
for some suitable boundary condition.
To achieve the target covariance $\Sigma(T)=\Sigma_T$, $\Pi, H$ must satisfy $\Pi(T)+H(T)=\Sigma_T$. Thus, we arrive at the coupled  system of differential equations
	\begin{subequations}\label{eq:gamecov}
	\begin{eqnarray}
	\dot \Pi \!\!&=&\!\! -A'\Pi- \Pi A +\Pi (B_1B_1'-B_2B_2')\Pi -Q
	\\
	\dot H \!\!&=&\!\! -A'H-HA-H(B_1B_1'-B_2B_2')H+Q
	\\&&+(\Pi+H)(B_1B_1'-B_2B_2'-CC')(\Pi+H) \nonumber
	\\ \Sigma_0^{-1} \!\!&=&\!\! \Pi(0)+H(0),
	\\ \Sigma_T^{-1} \!\!&=&\!\! \Pi(T)+H(T)
	\end{eqnarray}
	\end{subequations}
	
\begin{thm}
Suppose \eqref{eq:gamecov} has a solution over $t\in [0, T]$, then $F=\Pi(T)$ solves Problem \ref{pro:main}.
\end{thm}
{\em Proof:}
When the terminal cost in \eqref{eq:lqgame} is $x(T)'Fx(T)$, by Theorem \ref{thm:lqgame}, the differential game \eqref{eq:lqgame} has a unique solution given by \eqref{eq:nash}. The resulting state covariance satisfies the Lyapunov equation \eqref{eq:lyap}. Since by definition $\Sigma^{-1} = \Pi + H$, it matches the boundary condition $\Sigma(T) = \Sigma_T$. This completes the proof. $\Box$

The equation system \eqref{eq:gamecov} has a similar structure to that in \eqref{eq:sb}. When $B_1 = B, B_2=0$, \eqref{eq:gamecov} reduces to \eqref{eq:sb}. Indeed, in this case, player $2$ does not affect the system and Problem \ref{pro:main} boils down to the standard covariance control problem described in Section \ref{sec:cov}. When $B_1B_1'-B_2B_2'=CC'$, the two variables $\Pi, H$ are coupled only through the boundary conditions. If in addition $B_1B_1'-B_2B_2'\ge 0$ and $(A, B_1B_1'-B_2B_2')$ is controllable, then \eqref{eq:gamecov} has a closed form solution \cite{CheGeoPav14c}. In general, whether the solution to \eqref{eq:gamecov} exists remains an open question. However, just like the covariance control problem, the solution to Problem \ref{pro:main} can be approximated to an arbitrary precision, as shown below. 

\subsection{Alternative formulation}
Since the optimal policies to linear quadratic differential games are linear, without loss of generality, when taking an optimization approach, we can restrict ourself to the linear policies $u=K_1 x, v= K_2 x$. The cost function function $J_1(u,v)$, excluding the terminal cost, becomes
	\[
		 \int_0^T \tr(Q\Sigma +K_1\Sigma K_1'-K_2\Sigma K_2') dt.
	\]
This leads to the minimax problem
	\begin{eqnarray*}
	&&\hspace{-0.2cm}\min_{K_1} \max_{K_2, \Sigma>0} \int_0^T \tr(Q\Sigma +K_1\Sigma K_1'-K_2\Sigma K_2') dt
	\\&& \dot \Sigma = (A+B_1K_1+B_2 K_2)\Sigma 
	\\&&+\Sigma (A+B_1K_1+B_2K_2)' + CC'
	\\&& \Sigma(0) = \Sigma_0,\quad \Sigma(T) = \Sigma_T.
	\end{eqnarray*}
Adopting a standard reparametrization $Y_1 = K_1 \Sigma, Y_2 = K_2\Sigma$, we obtain
	\begin{eqnarray*}
	&&\hspace{-0.3cm}\min_{Y_1} \max_{Y_2, \Sigma>0} \int_0^T \tr(Q\Sigma +Y_1\Sigma^{-1} Y_1'-Y_2\Sigma^{-1} Y_2') dt
	\\&&\hspace{-0.3cm} \dot \Sigma = A\Sigma +\Sigma A' + B_1Y_1+Y_1'B_1'+B_2Y_2+Y_2'B_2'+CC'
	\\&& \hspace{-0.3cm}\Sigma(0) = \Sigma_0,\quad \Sigma(T) = \Sigma_T.
	\end{eqnarray*}
Invoking the Schur complement, we deduce the equivalent problem
	\begin{subequations}\label{eq:SDP}
	\begin{eqnarray}
	&&\hspace{-0.5cm}\min_{Y_1,Z_1} \max_{Y_2, Z_2, \Sigma} \int_0^T \tr(Q\Sigma +Z_1-Z_2) dt \label{eq:SDP1}
	\\&&\hspace{-0.5cm} \dot \Sigma \!=\! A\Sigma \!+\!\Sigma A' \!+\! B_1Y_1\!+\!Y_1'B_1'\!+\!B_2Y_2\!+\!Y_2'B_2'\!+\!CC' \label{eq:SDP2}
	\\&&\hspace{-0.5cm} \Sigma(0) = \Sigma_0,\quad \Sigma(T) = \Sigma_T \label{eq:SDP3}
	\\&&\hspace{-0.5cm} \left[\begin{matrix}Z_1 & Y_1 \\Y_1' & \Sigma\end{matrix}\right] \ge 0, 
	\quad \left[\begin{matrix}Z_2 & Y_2 \\Y_2' & \Sigma\end{matrix}\right] \ge 0. \label{eq:SDP4}
	\end{eqnarray}
	\end{subequations}
The cost is convex over $Y_1, Z_1$ and concave over $Y_2, Z_2, \Sigma$, and the constraints are convex, therefore the above formulation \eqref{eq:SDP} is a convex-concave minimax problem. We remark that the feasible set is not empty \cite{CheGeoPav14b}. Though, the optimization variable is of infinite dimension. The problem in a finite-dimension setting has been extensively studied and many algorithms have been proposed \cite{Ber97,NedOzd09,CheLanOuy14}. The Lagrangian multiplier associated with the constraint \eqref{eq:SDP2} turns out to be $\Pi$. The optimal $F$ in Problem \ref{pro:main} can therefore be recovered by taking $F=\Pi(T)$.

\section{Infinite horizon cases}\label{sec:inf}
In this section we investigate the covariance steering problem for differential games in the infinite horizon setting. The goal is to select some incentive function so that the system-state attains a stationary Gaussian distribution with a specified covariance, formally stated below.
 \begin{problem}\label{pro:inf}
 Consider the differential game associated with the dynamics
	\begin{equation}\label{eq:dynamicsinf}
		dx(t) = Ax(t) dt + B_1 u(t) dt + B_2 v(t) dt + Cdw
	\end{equation}
and costs $J_2 = -J_1$, 
	\begin{equation}\label{eq:costinf}
		J_1 (u,\!v)\! =\! \limsup_{T\rightarrow \infty} \frac{1}{T} \mE\{\int_0^T [x(t)'Qx(t) \!+\! \|u(t)\|^2 \!-\! \|v(t)\|^2] dt\}
	\end{equation}
where $Q$ is a design variable. Find a $Q$ such that the state reaches the stationary distribution $\cN(0, \Sigma)$, provided the two players are rational. By ``rational'' we mean that they both seek to gain maximal advantage, while at the same time, assume that their opponent does the same.
 \end{problem}
In the above, we assume that $(A, C)$ is controllable, to avoid possibly degeneracy of the state distribution. 
 Again, without loss of generality, we only consider the linear policy $u=K_1 x, v = K_2 x$. Assuming the controlled system 
	\[
		dx = (A+B_1K_1+B_2 K_2) xdt+Cdw
	\]
is stable, namely, $A+B_1K_1+B_2 K_2$ is Hurwitz, then the state covariance reaches the specified value $\Sigma>0$ with
	\[
		(A+B_1K_1+B_2K_2)\Sigma+\Sigma(A+B_1K_1+B_2K_2)' + CC' = 0.
	\]
The cost $J_1$, with $Q=0$, can then be written as
	\[
		\tr(K_1 \Sigma K_1'-K_2 \Sigma K_2').
	\]

Consider the minimax optimization
	\begin{subequations}\label{eq:infSDP}
	\begin{eqnarray}
	&&\min_{K_1} \max_{K_2}~ \tr (K_1 \Sigma K_1'-K_2 \Sigma K_2')\label{eq:infSDP1}
	\\&&
	(A+B_1K_1+B_2K_2)\Sigma+\Sigma(A+B_1K_1+B_2K_2)'\nonumber
	\\&& + CC' = 0,\label{eq:infSDP2}
	\end{eqnarray}
	\end{subequations}
where $\Sigma>0$ is a given target covariance. Clearly, the cost is convex over $K_1$ and concave over $K_2$. Suppose the feasible set is not empty, then a solution to \eqref{eq:infSDP} exists. In fact, under the assumption that $B_1, B_2$ are of full column rank, the solution is unique. With the new parametrization $Y_1 = K_1 \Sigma, Y_2 = K_2\Sigma$, \eqref{eq:infSDP} can be rewritten as
	\begin{subequations}\label{eq:infY}
	\begin{eqnarray}
	&&\min_{Y_1} \max_{Y_2}~ \tr (Q\Sigma + Y_1 \Sigma^{-1} Y_1'-Y_2 \Sigma^{-1} Y_2') \label{eq:infY1}
	\\&&
	A\Sigma+\Sigma A'+B_1Y_1+Y_1'B_1'+B_2Y_2+Y_2'B_2'\nonumber
	\\&& + CC' = 0. \label{eq:infY2}
	\end{eqnarray}
	\end{subequations}
The feasible set is not empty if and only if \eqref{eq:infY2} has a solution $(Y_1, Y_2)$, which is equivalent to the condition \cite{Geo02}
	\begin{equation}
	{\rm rank}\left[\begin{matrix}A\Sigma+\Sigma A'+CC' & B\\B' & 0\end{matrix}\right]
	=
	{\rm rank}\left[\begin{matrix}0 & B\\B' & 0\end{matrix}\right],
	\end{equation}
where $B=[B_1\, B_2]$. 

When $\range(B)\subset\range(C)$, the constraint \eqref{eq:infSDP2} guarantees that $A+B_1K_1+B_2K_2$ is Hurwitz, by Lyapunov theory. In general, the condition that $A+B_1K_1+B_2K_2$ be Hurwitz needs to be verified separately; it may have eigenvalues on the imaginary axis. Nevertheless, it is possible to maintain a state covariance that is arbitrarily close. Indeed, for any feasible $K_1, K_2$, let
	\[
		K_1^\epsilon = K_1 - \frac{1}{2}\epsilon B_1'\Sigma^{-1},\quad K_2^\epsilon = K_2- \frac{1}{2}\epsilon B_2'\Sigma^{-1}
	\]
for $\epsilon>0$, then
	\begin{eqnarray*}
		&&(A+B_1K_1^\epsilon+B_2K_2^\epsilon)\Sigma+\Sigma(A+B_1K_1^\epsilon+B_2K_2^\epsilon)'
		\\ &=& -CC'-\epsilon BB'
		\le -\epsilon BB',
	\end{eqnarray*}
which implies that $A+B_1K_1^\epsilon+B_2K_2^\epsilon$ is Hurwitz. The real state covariance $\Sigma_\epsilon$ satisfies
	\[
		(A+B_1K_1^\epsilon+B_2K_2^\epsilon)\Sigma_\epsilon+\Sigma_\epsilon(A+B_1K_1^\epsilon+B_2K_2^\epsilon)'+CC' =0,
	\]
Its difference $\Delta = \Sigma-\Sigma_\epsilon\ge0$ to $\Sigma$ solves 
	\[
		(A+B_1K_1^\epsilon+B_2K_2^\epsilon)\Delta+\Delta(A+B_1K_1^\epsilon+B_2K_2^\epsilon)' = -\epsilon BB',
	\]
which is clearly of order $o(\epsilon)$.

Therefore, without loss of generality, we assume that the unique solution to \eqref{eq:SDP} corresponds to a stable closed loop system. Next we discuss how \eqref{eq:infSDP} is related to Problem \ref{pro:inf}. The minimax problem \eqref{eq:SDP} can be rewritten as
	\begin{subequations}\label{eq:infref}
	\begin{eqnarray}
	&&\min_{K_1} \max_{K_2,\hat\Sigma}~ \tr (K_1 \Sigma K_1'-K_2 \Sigma K_2')
	\\&&
	(A+B_1K_1+B_2K_2)\hat\Sigma+\hat\Sigma(A+B_1K_1+B_2K_2)'\nonumber
	\\&& + CC' = 0,
	\\&&
	\hat\Sigma = \Sigma.\label{eq:infref3}
	\end{eqnarray}
	\end{subequations}
Relaxing the last equality constraint using Lagrange multiplier method we arrive at
	\begin{eqnarray*}
	&&\min_{K_1} \max_{K_2,\hat\Sigma}~ \tr (K_1 \Sigma K_1'-K_2 \Sigma K_2') + \tr(\Pi \hat\Sigma)
	\\&&
	(A+B_1K_1+B_2K_2)\hat\Sigma+\hat\Sigma(A+B_1K_1+B_2K_2)'
	\\&& + CC' = 0,
	\end{eqnarray*}
which is exactly the optimization formulation to a standard infinite horizon differential game problem with dynamics \eqref{eq:dynamicsinf} and cost
	\[
		\limsup_{T\rightarrow \infty} \frac{1}{T} \mE\{\int_0^T [x(t)'\Pi x(t) + \|u(t)\|^2 - \|v(t)\|^2] dt\}.
	\]
In view of \eqref{eq:costinf} we conclude that the solution to Problem \ref{pro:inf} is $Q=\Pi$ where $\Pi$ is the optimal Lagrangian multiplier of \eqref{eq:infref} associated with the constraint \eqref{eq:infref3}.


\section{Conclusion}\label{sec:con}
We formulated and studied a class of two-player zero-sum linear-quadratic differential games with the added specification of a terminal state covariance of the combined system dynamics. Such an added specification may be used to limit the range of operation for the combined two-player dynamics. Thus, the two players must abide by this extra specification while, independently, also compete to ensure optimal individual gains. We show that a suitable modification of the cost functional provides the appropriate incentive that drives the combined dynamics towards meeting the terminal state constraint. We characterized solutions and numerical algorithms that effect convex-concave minimax optimization. 

The present work is perhaps the first attempt to extend covariance control to the differential game setting. Potential applications range from probabilistic path planing involving competitors, to the classic pursuit-evasion problems. 
Possible future directions include general distribution-steering for more general dynamics, nonzero-sum games, and games involving more than two players. On the technical side, it will be important to study the implications of relaxing the standing assumption made in Section \ref{sec:LQDG} that the pairs $(A,\, B_1)$ and $(A,\, B_2)$ are controllable, to controllability of the pair $(A,\,[B_1,\,B_2])$. Under our current assumption, individual players have control over the complete combined state space, whereas the relaxed condition will allow for the possibility that players can only control respective individual dynamics.

{
\bibliographystyle{IEEEtran}
\bibliography{./refs}
}
\end{document}